# Sampling unknown large networks restricted by low sampling rates

Bo Jiao

Email: jiaoboleetc@outlook.com

*School of Information Science & Technology, Xiamen University Tan Kah Kee College,*

*Fujian, 363123, China*

**Abstract:** Graph sampling plays an important role in data mining for large networks. Specifically, larger networks often correspond to lower sampling rates. Under the situation, traditional traversal-based samplings for large networks usually have an excessive preference for densely-connected network core nodes. Aim at this issue, this paper proposes a sampling method for unknown networks at low sampling rates, called SLSR, which first adopts a random node sampling to evaluate a degree threshold, utilized to distinguish the core from periphery, and the average degree in unknown networks, and then runs a double-layer sampling strategy on the core and periphery. SLSR is simple that results in a high time efficiency, but experiments verify that the proposed method can accurately preserve many critical structures of unknown large scale-free networks with low sampling rates and low variances.

## 1 Introduction

Graph sampling extracts nodes or edges to create subgraphs representing an original network, which is often used as pre-processing before data mining to reduce the scale of datasets [1-5] or as post-processing after optimization to represent the original network more accurately [6-9]. The former usually adopts traversal-based samplings [5] that simulate walkers travelling on unknown networks based on the neighbor information of the nodes they are accessing. The traversal-based samplings [5] are time efficient, which enables complex mining algorithms, such as graph convolutional networks [1], subgraph pattern mining [2,3] and network embedding [4], to be applied to networks with more than one million nodes. The latter adopts samplings that construct optimization models [6-9] to minimize the difference between known original networks and sampled subgraphs. Approximation algorithms used for solving the optimization models are powerful in representing objective structures of the original networks but are time-consuming [6-9]. This paper focuses on the traversal-based samplings in pre-processing systems that travel on unknown original networks, and intents to represent more important structures of the networks with high time efficiency.

Metropolis-hastings random walk (MHRW) [10] and simple random walk (SRW) [10,11] are two classical traversal-based samplings. Based on Markov chain random models, MHRW is unbiased, which samples each node with uniform stationary distribution, whereas, SRW is biased, that is, the probability of a node being sampled is proportional to its degree [10,11]. This paper focuses on ubiquitous scale-free networks exhibiting core-periphery structures, which consist of a dense core and a sparse periphery [12-15]. The core determines many important structures of the networks, such as low-diameter; however, it is almost ignored by unbiased samplings since the number of nodes in the core is extremely small. On the contrary, biased samplings preserve more structures determined by the core nodes that have high-degrees. However, the above-mentioned node sampling probability of SRW corresponds to the convergence state of Markov chain [11].

Sampling rate is defined as the ratio of the number of nodes (or edges) between sampled and original networks [5]. Low sampling rates are needed to improve time efficiency on large networks, but also make it difficult to achieve the convergence of Markov chain. In the core-periphery structures of scale-free networks, each core node is well-connected by periphery nodes but the latter are not well-connected to each other [12-15], that is, the walkers of biased samplings are more likely to be attracted to the core under constraints of low sampling rates, resulting in loss of structures related to the periphery that occupies the vast majority of nodes in the networks. Thus, this paper proposes a sampling for unknown networks at low sampling rates, called SLSR, objective to achieve a balanced sampling on the core-periphery structures.

The organization of this paper is as follows: Section 3 investigates the problem formulation and design principles of SLSR. Section 4 provides a random node sampling to evaluate the average degree (AD) and degree threshold (DT) of original unknown large networks. Section 5 designs the traversal-based sampling SLSR. Specifically, SLSR starts by the AD and DT evaluation, then limits the sampling process to the periphery using the DT, and designs a bisection method constrained by the AD to preserve the core structure. Section 6 evaluates SLSR with related methods and verifies that SLSR can capture many critical structures except for degree, including shortest path length, clustering, graph spectrum, centrality and communities.

The contributions of this paper are as follows:
- Analysis of the advantages of the random node sampling in capturing the DT and its high time efficiency, as well as the shortcomings of the sampling in subgraph representation, such as, the loss of critical high-degree core nodes and periphery topological structure.
- Designing a simple traversal-based sampling that only relies on node set and the adjacent node information of sampled nodes, without involving complex topological characteristics, but can preserve critical properties at low sampling rates. Simplicity corresponds to a high time efficiency that is important for pre-processing systems.
- Analysis of entropy and variance of sampled subgraphs at low sampling rates. Assuming $\Omega$ is a sample space with $n$ samples, then the probability of extracting a sample from $\Omega$ uniformly at random is $1/n$, and the entropy of the probability distribution is $H = log n = -\sum \frac{1}{n} log \frac{1}{n}$. That is, with increasing $n$, the uncertainty measured by the entropy grows [16]. The traversal-based samplings randomly choose next node $w$ from the adjacent nodes of a current node $v$, that is, the adjacent node set of the node $v$ constructs a sample space for randomly choosing $w$. We prove that SLSR sharply reduces the scale of the sample space at most cases by a simple and deterministic bisection method (i.e., reduces the entropy), and experimentally verify that the reduced entropy can ensure the low variances of many critical statistics of the sampled SLSR subgraphs.
- We experimentally obtain that, the smaller and denser the cores in original networks, the stronger the preference of the traditional traversal-based samplings for high-degree core nodes at low sampling rates, which is difficult to be mathematically proven by the Markov chain theory [10], because the low sampling rates prevent the Markov chain random process from reaching convergence state.
- Time efficiency and community visualization are analyzed in depth.
- The codes of SLSR are provided at https://github.com/jiaoboleetc/SLSR.

## 2 Related work

### 2.1 Graph sampling on unknown networks

**Node/edge-based samplings** choose a set of nodes (or edges) at random and extract the subgraphs induced by the chosen nodes (or edges), including uniform samplings, such as random node (RN) and random edge (RE), and non-uniform samplings, such as random degree node (RDN) and random PageRank node (PRN) [5,17]. Specifically, nodes can be sampled proportional to the degree centrality by RDN [5], and proportional to the PageRank weight by PRN [5]. Recently, Wang et al. [18] investigated the relations between edges and their edge neighbors caused by the reconcile of scale-free and self-similarity, and proposed a series of sampling algorithms based on the relations, which can keep important statistical characteristics of original networks.

**Traversal-based samplings** start with one or more seeds and crawl on unknown original networks based on the neighbor information of the nodes they are accessing. Forest fire (FF), which is a variant of breadth first (BF) and snow ball (SB), performs superior in time efficiency since each node in the unknown networks is traversed no more than once [19]. FF starts by a random seed, then burns a fraction of its neighbors that have not been traversed, where the fraction is randomly drawn from a geometric distribution, and the process is recursively repeated for each burnt neighbor until the desired sample size is obtained [5]. SRW [11] starts by a random seed, and moves from a node to one of its neighbors chosen uniformly at random, until the expected fraction of nodes is collected. In addition, more random walk samplings, namely, non-backtracking random walk (NBRW), circulated neighbor random walk (CNRW), and common neighbor awareness random walk (CNARW) [20], have been proposed to reduce the asymptotic variance of sampled subgraphs and overcome the slow convergence of SRW, as simple random walker tends to be stuck in local loops [20]. The principles and pseudo codes of the three improved random walk samplings can be obtained in the recent review articles [10,20]. Rank degree (RD) is a multi-seed sampling [21], which adopts a predetermined number of random starting seeds to avoid the sampling trapped locally, then iteratively explores top-$k$ highest-degree neighbors of each seed and adds them to the seed set. Moreover, some samplings were designed to capture specific network structures, such as, community structure expansion (CSE) [22]. Recently, the node/edge-based and traversal-based samplings have become important tools for efficient network intervention and AD evaluation on large unknown networks [23].

**Stream-based samplings** generate subgraphs from activity networks that can be treated as a stream of edges [24,25]. In the networks, besides the unknown topology, the node set and the neighbor information of any node are unobtainable.

### 2.2 Graph sampling on known networks

If all network information of a dataset is known, complex structures hidden in the dataset can be discovered in advance. Hong et al. [6] first extracted precise structures, such as $k$-core, closeness, betweenness, and eigenvector centrality, from known original networks, and then reduced the scale of the networks under the guidance of the structures. Martin et al. [8] created an optimization model for large-scale network reduction towards scale-free structure. Jiao et al. [9] adopted a strategy of removing edges from known original networks one by one. However, a lower sampling rate means more edges need to be removed. Sampling on known networks helps preserve more precise structures, but usually comes at the cost of time [6-9].

## 3 Problem formulation and design principles

## 3.1 Problem formulation

This paper focuses on simple, undirected, and scale-free original networks, in which self-loops, multi-edges, and direction of edges are ignored. We assume that the topological information of the original networks, such as, community, clique, and global statistical characteristics, is unknown. But we assume that the node set and the neighbors of sampled nodes can be accessed [5,17-23]. We intend to quickly obtain subgraphs representing the unknown large original networks. The notions used by our SLSR sampling are listed in Table 1.

Table 1. Notions and descriptions.

| Notions | Descriptions |
|---|---|
| $G_{org} = (V_{org}, E_{org})$ | An unknown original network $G_{org}$ where $V_{org}$ and $E_{org}$ respectively denote the node set and edge set. Please note that $G_{org}$ is a simple and undirected graph. |
| $\bar{d}_{org}$ | The average degree of $G_{org}$ that can be evaluated by a random node sampling [23]. |
| $\ddot{d}_{org}$ | A degree threshold of $G_{org}$ that can be evaluated by a random node sampling. |
| $N_{org}(v)$ | The set of neighbor nodes of a sampled node $v$ in $G_{org}$, which can be obtained by the traversal-based samplings. |
| $\|\cdot\|$ | The cardinality of a set. |
| $d_{org}(v)$ | The degree of a sampled node $v$ in $G_{org}$, where $d_{org}(v) = \|N_{org}(v)\|$. |
| $S_1 - S_2$ | A set consisting of elements that belong to $S_1$ but not to $S_2$, where $S_1$ and $S_2$ denote two sets. |
| $R_{RN}$ | A sampling rate of the random node sampling for evaluating AD and DT. |
| $R_{SLSR}$ | A sampling rate of the SLSR sampling. |
| $G_{sub} = (V_{sub}, E_{sub})$ | A sampled subgraph of $G_{org}$ where $V_{sub} \subseteq V_{org}$ and $E_{sub} \subseteq E_{org}$. |

## 3.2 Design principles

The classical Barabasi-Albert (BA) scale-free evolving network model [26] confirms that the degree distribution of the network almost remains the same as the scale changes. Please note that the distribution only represents low-degrees of nodes in periphery, ignoring high-degrees of nodes in core, because the number of core nodes is extremely smaller than that of periphery nodes [12-15]. Based on the preferential attachment (PA) rule adopted by the BA model [26], which attaches each newly-added node preferentially to high-degree nodes, the degrees of core nodes quickly grow with increasing network scale, that is, the larger an original network, the greater the difference in degree between the core and periphery nodes, which causes the biased samplings on large networks to be overly attracted to the core at low sampling rates. The degree distribution is an important metric and the existing biased samplings are good at capturing the metric under specific conditions [17-22]. Thus, the first principle **P1** is to create a core-periphery framework in which the existing biased samplings continue to be used but are only limited to the periphery sampling, that is, the core, which hinders the capture of the degree distribution, is stripped off and processed separately.

During changes in a scale-free network, such as scale-reduction, the core has a low variability [15]. In addition, based on the fractal characteristic [26], the communities of the network also represent core-periphery structures [9,27]. Specifically, the community cores are mainly located in the core of the network, that is, the network core represents the structure of community centers. Thus, the second principle **P2** is to maximize the preservation of the connections in the network core.

Owing to sparse connections between periphery nodes, faster information exchange between these nodes depends on core nodes [12-15]. Specifically, based on the PA rule [26], the higher the degree of a core node, the greater its probability of being connected by other newly-added periphery nodes, that is, the core node has stronger ability to shorten the path length between periphery nodes. Thus, the third principle **P3** is to preserve a proportion of core neighbors with top highest degrees for each sampled periphery node, where the proportion can be determined by the AD of original networks that can be evaluated by a random node sampling [23]. Please note that the second and third principles are helpful in preserving the path length distribution.

The periphery is the main contributor of the clustering coefficient distribution since it occupies the vast majority of nodes in the network [12-15]. Based on the PA rule [26], the neighbors of a periphery node tend to be located in the core, that is, the connections between high-degree core nodes has a significant impact on the distribution. Thus, the above-mentioned three principles marked as **P1**, **P2**, and **P3** are helpful in preserving the distribution.

## 4 Random node sampling for evaluating parameters

### 4.1 Random node evaluation

Core-periphery detection [27] refers to a partition of a network into two groups of nodes called core and periphery, which is a useful tool to realize P1. An important procedure of the detection is to provide rank orders of nodes for the partition. Many measures based on clique, community, centrality, and probability [12-15,27], have been adopted for the rank. These complex measures can help improve detection accuracy, but are difficult to be quickly evaluated on unknown networks. Thus, we use a simple measure, namely degree, to rank the nodes, and adopt a random node sampling to evaluate the AD and DT of the unknown networks that are critical for P1 and P3. Specifically, nodes with degrees larger than DT are classified to the core, while other nodes are divided to the periphery. To clearly distinguish the core nodes and periphery nodes, the DT is determined by maximizing the number of edges connecting the two types of nodes.

---

**Random node sampling algorithm: Evaluating the AD and DT**

---

1: **Input:** A sampling rate $R_{RN}$, and an original network $G_{org} = (V_{org}, E_{org})$ with unknown topological information, but the node set and the neighbors of sampled nodes in the network can be accessed.

2: **Output:** $\bar{d}_{org}$, namely the evaluated AD, and $\ddot{d}_{org}$, namely the evaluated DT, of $G_{org}$.

3: Randomly extract $\|V_{org}\| \times R_{RN}$ nodes from $V_{org}$ with uniform distribution to form a node set $V$, and derive $d_{max} = \max\{d_{org}(v)|v \in V\}$.

4: Derive $\bar{d}_{org} \leftarrow \sum_{v \in V} d_{org}(v) / \|V\|$.       %Please note that $d_{org}(v) = \|N_{org}(v)\|$.

   %Please note that $N_{org}(v)$ is the set of neighbor nodes of node $v$ in the original network $G_{org}$, not in the subgraph induced by the node set $V$.

5: Initialize $vertical\_edge\_set \leftarrow \emptyset$ that consists of vertical edges defined as the edges connected by a periphery node and a core node.

6: Initialize $k \leftarrow 1$, and $\ddot{d}_{org} \leftarrow 0$.

7: **While** $k \leq d_{max}$ **do**

8:    Update $\ddot{d}_{org} \leftarrow k$, and derive a node set $V_k = \{v | v \in V \wedge d_{org}(v) = \ddot{d}_{org}\}$.

9:    Derive two sets caused by the update of $\ddot{d}_{org}$ from $k-1$ to $k$, namely, $added\_edge\_set$, which consists of edges that are newly added to $vertical\_edge\_set$, and $removed\_edge\_set$, which consists of edges that should be removed from $vertical\_edge\_set$.

$$added\_edge\_set = \{(v, u) | v \in V_k \wedge u \in N_{org}(v) \wedge d_{org}(u) > \ddot{d}_{org}\}$$

$$removed\_edge\_set = \{(v, u) | (v, u) \in vertical\_edge\_set \wedge d_{org}(v) < \ddot{d}_{org} \wedge d_{org}(u) = \ddot{d}_{org}\}$$

   Update $vertical\_edge\_set \leftarrow vertical\_edge\_set \cup added\_edge\_set - removed\_edge\_set$.

10:   **If** $\|added\_edge\_set\| \leq \|removed\_edge\_set\|$ **do**   Update $\ddot{d}_{org} \leftarrow k - 1$;   **Break;**   **End If**

11:   Update $k \leftarrow k + 1$.

12: **End While**                     %Lines 4 and 10 respectively output the evaluated AD and DT.

---

In the above sampling, line 4 evaluates the AD of the unknown original network $G_{org}$, and lines 5 to 12 evaluate the DT of $G_{org}$. Our random node sampling is different from the RN sampling [5] introduced in Section 2.1 that generates a subgraph induced by randomly chosen nodes. Our sampling collects the degrees $d_{org}(v)$ and neighbors $N_{org}(v)$ of the randomly chosen nodes in $G_{org}$, and evaluates the AD and DT based on the degrees and neighbors.

## 4.2 Analysis of the AD evaluation

Recently, Qi et al. [23] confirmed the effectiveness of the random node sampling on the evaluation of AD. We further analyze the shortcomings of the sampling in subgraph representation.

**Table 2**. Degree rank and the number of nodes with a certain degree in a scale-free network. (com-Youtube with 1,134,879 nodes [28], described in Section 6.2, was chosen for the analysis.)

| Rank $i$ | The $i^{th}$ highest-degree $k$ | The number of nodes with degree $k$ | Rank $i$ | The $i^{th}$ highest-degree $k$ | The number of nodes with degree $k$ |
|---|---|---|---|---|---|
| 1 | 28,754 | 1 | 800 | 180 | 17 |
| 2 | 14,641 | 1 | 900 | 80 | 121 |
| 3 | 11,281 | 1 | 978 | 2 | 182,237 |
| 500 | 490 | 2 | 979 | 1 | 602,530 |

All degrees existing in the com-Youtube network [28] are ranked by decreasing order in Table 2, which shows that the top highest-degrees are related to only one or several nodes. Please note that the phenomenon in Table 2 is common in scale-free networks. With uniform distribution, the probability of a $k$-degree node being sampled is equal to $P(k)$ that is defined as the ratio of the number of $k$-degree nodes to the total number of nodes in the original network. Thus, the top highest-degree nodes in scale-free networks are easily lost by unbiased samplings at low sampling rates, such as the random node sampling, which induces that the AD of the subgraph induced by the nodes randomly chosen by the RN sampling [5] is extremely small, as shown in Fig. 1.

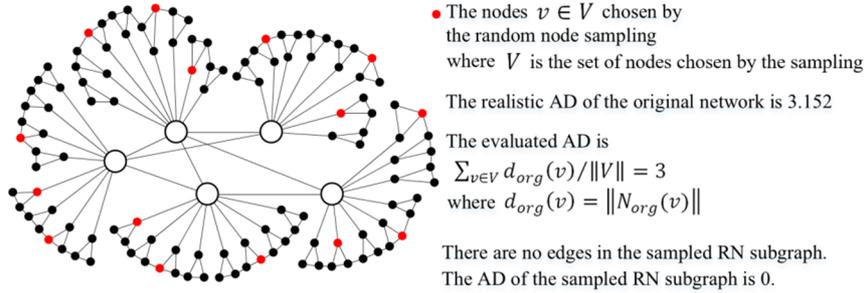

**Fig. 1.** Random node sampling on a simple core-periphery graph, in which $N_{org}(v)$ is defined in the original network, not in the RN subgraph, and can be obtained by the traversal-based samplings. The random node sampling is suitable for the evaluation of the AD of the original network [23], but cannot directly output a sampled subgraph capturing the degree property.

The AD of the original network $G_{org}$ is equal to $\sum kP(k)$, where $k$ is the degree defined in $G_{org}$. The random node sampling is prone to losing top highest-degrees $d$. However, owing to the extremely small $P(d)$, the loss of these degrees has almost no impact on the AD evaluation.

## 4.3 Analysis of the DT evaluation

Let us return to the random node sampling algorithm described in Section 4.1. Assuming that $\ddot{d}_{org}$ in line 8 is not more than the actual DT value, then the $V_k$ nodes in line 8 and the nodes $u$ with $d_{org}(u) = \ddot{d}_{org}$ in $removed\_edge\_set$ (in line 9) are classified into the periphery, that is,

$$\|added\_edge\_set\| > \|removed\_edge\_set\|, \qquad (1)$$

because $added\_edge\_set$ (in line 9) contains the edges that connect the peripheral $V_k$ nodes to core nodes in the original network $G_{org}$, and $removed\_edge\_set$ consists of the edges connecting two periphery nodes. Please note that there are dense connections between core and periphery nodes but sparse connections between periphery nodes [12-15,27]. Although the node set $V$ in line 3 (chosen by the random node sampling) losses top highest degree nodes in the core, most of nodes in $V$ can directly reach the core through only one jump in the scale-free original network $G_{org}$, because the PA rule [26] causes the core to be densely connected by the periphery [12-15,27], which ensures that the top highest degree nodes in $G_{org}$ are not lost in the neighbor sets $N_{org}(v)$ for most of nodes $v$ that falls in the node set $V$ (the result will be further verified in Section 4.5). Thus, all the connections between the peripheral $V_k$ nodes and the core nodes in $G_{org}$ are preserved in the edge set $added\_edge\_set = \{(v,u)|v \in V_k \wedge u \in N_{org}(v) \wedge d_{org}(u) > \ddot{d}_{org}\}$ in line 9.

The nodes in the original network $G_{org}$ are independently chosen with uniform distribution; thus, the random node sampling not only loses top highest degree nodes in the core but also ignores the complex topological correlation between sampled periphery nodes. However, the determination of the DT value depends on the connections between core and periphery nodes, while it is weakly correlated with the connections between periphery nodes. The uniform distribution enables the random node sampling to choose the periphery nodes without preference, which is critical for the accurate DT evaluation of our random node sampling.

### 4.4 Analysis of the variance and runtime

We choose five real-world large scale-free networks [28], described in Section 6.2, for the variance and runtime analysis, as listed in Table 3.

**Table 3**. Mean and standard errors (below the mean) of the evaluated AD, evaluated DT and runtime (seconds) from 100 independent realizations for each sampling rate $R_{RN}$. The real AD and DT were obtained by $R_{RN} = 100\%$. The running environment is illustrated in Section 6.3.5.

| Random node sampling (designed in Section 4.1) on original real-world networks [28] | | Sampling rate $R_{RN}$ | | | | | |
|---|---|---|---|---|---|---|---|
| | | 15% | 20% | 25% | 30% | 35% | 100% |
| ego-Twitter | Evaluated AD $\bar{d}_{org}$ | 33.01 ±0.521 | 32.93 ±0.450 | 33.01 ±0.413 | 33.02 ±0.363 | 32.96 ±0.349 | 33.01 ±0 |
| | Evaluated DT $\ddot{d}_{org}$ | 68.18 ±4.085 | 68.34 ±3.687 | 68.48 ±3.666 | 70.00 ±3.552 | 70.65 ±3.561 | 76.00 ±0 |
| | Runtime (Seconds) | 0.309 ±0.086 | 0.465 ±0.103 | 0.587 ±0.143 | 0.744 ±0.222 | 0.797 ±0.243 | — |
| loc-Gowalla | Evaluated AD $\bar{d}_{org}$ | 9.676 ±0.292 | 9.687 ±0.212 | 9.662 ±0.206 | 9.654 ±0.184 | 9.661 ±0.174 | 9.662 ±0 |
| | Evaluated DT $\ddot{d}_{org}$ | 27.35 ±2.921 | 27.71 ±2.768 | 28.48 ±2.664 | 29.28 ±2.574 | 29.47 ±1.951 | 32.00 ±0 |
| | Runtime (Seconds) | 0.581 ±0.066 | 0.797 ±0.102 | 1.065 ±0.106 | 1.218 ±0.080 | 1.408 ±0.085 | — |
| com-DBLP | Evaluated AD $\bar{d}_{org}$ | 6.629 ±0.046 | 6.621 ±0.032 | 6.619 ±0.029 | 6.623 ±0.028 | 6.618 ±0.021 | 6.621 ±0 |
| | Evaluated DT $\ddot{d}_{org}$ | 10.05 ±0.297 | 10.02 ±0.200 | 10.02 ±0.245 | 10.01 ±0.173 | 10.01 ±0.100 | 10.00 ±0 |
| | Runtime (Seconds) | 0.636 ±0.238 | 0.912 ±0.281 | 0.982 ±0.335 | 1.249 ±0.491 | 1.350 ±0.444 | — |
| web-Stanford | Evaluated AD $\bar{d}_{org}$ | 14.18 ±0.695 | 14.09 ±0.526 | 14.10 ±0.572 | 14.08 ±0.439 | 14.14 ±0.423 | 14.14 ±0 |
| | Evaluated DT $\ddot{d}_{org}$ | 37.69 ±0.597 | 37.68 ±0.601 | 37.72 ±0.514 | 37.68 ±0.468 | 37.74 ±0.440 | 38.00 ±0 |
| | Runtime (Seconds) | 0.701 ±0.253 | 0.939 ±0.327 | 1.196 ±0.410 | 1.429 ±0.470 | 1.759 ±0.589 | — |
| com-Youtube | Evaluated AD $\bar{d}_{org}$ | 5.277 ±0.113 | 5.251 ±0.086 | 5.259 ±0.083 | 5.271 ±0.070 | 5.274 ±0.067 | 5.265 ±0 |
| | Evaluated DT $\ddot{d}_{org}$ | 34.49 ±2.783 | 35.44 ±2.388 | 35.67 ±2.165 | 35.81 ±2.218 | 36.09 ±2.020 | 39.00 ±0 |
| | Runtime (Seconds) | 2.401 ±0.714 | 3.273 ±0.924 | 4.360 ±0.965 | 5.094 ±0.922 | 6.560 ±1.038 | — |

Based on Table 3, a low sampling rate $R_{RN}$ is capable of evaluating the AD and DT values. However, with increasing $R_{RN}$, the standard errors of the evaluated values show a decreasing trend. Owing to the very high time efficiency of the random node sampling, which is induced by that the sampling ignores the complex topological correlation among the randomly chosen nodes, we choose $R_{RN} = 35\%$ to pursue a low variance. Note that the random node sampling cannot directly output a subgraph capturing the degree properties or other important properties, as shown in Fig. 1; however, our purpose is to obtain a subgraph representing the original network.

### 4.5 Analysis of diverse core-periphery structures

Based on the DT value $\ddot{d}_{org}$, the original network $G_{org} = (V_{org}, E_{org})$ can be partitioned into core nodes with $d_{org}(v) > \ddot{d}_{org}$, periphery nodes with $d_{org}(v) \leq \ddot{d}_{org}$, core edges that connect two core nodes, periphery edges that connect two periphery nodes, and vertical edges that connect a periphery node to a core node. As shown in Table 4, the five real-world scale-free networks consist of a dense core and a sparse periphery. In addition, a few core nodes are densely connected by the periphery, and more than 55% of periphery nodes can directly reach the core through only one jump. Moreover, for the com-Youtube, web-Stanford and loc-Gowalla networks in Table 4, we find that their core node percentages are much smaller and their core edge densities (defined as the ratio of the number of core edges to the number of core nodes) are much denser than those of other networks. Restricted by low sampling rates, the smaller and denser the core, the stronger its attraction to traditional traversal-based samplings. Because the Markov chain theory cannot achieve convergence at low sampling rates [20], this paper will further confirm the impact of the core structure of the original networks on the sampling results through experimental comparisons in Section 6.

**Table 4**. Percentage distributions of nodes and edges in the core-periphery structures partitioned by $\ddot{d}_{org}$. Two DT values were chosen for each original network: bold represents the accurate value with $R_{RN} = 100\%$, and non-bold represents the mean of the evaluated values with $R_{RN} = 35\%$, as shown in Table 3. $R_{per}$ is defined as the ratio of the number of periphery nodes that can directly reach core through only one jump to the total number of periphery nodes.

| Original networks [28] | $\ddot{d}_{org}$ | Node percentage distribution | | Edge percentage distribution | | | $R_{per}$ |
|---|---|---|---|---|---|---|---|
| | | Core | Periphery | Core | Vertical edges | Periphery | |
| ego-Twitter | **76** | **9.88%** | **90.12%** | **25.37%** | **46.79%** | **27.83%** | **97.83%** |
| | 70.65 | 11.22% | 88.78% | 28.47% | 46.55% | 24.98% | 98.21% |
| loc-Gowalla | **32** | **5.31%** | **94.69%** | **26.09%** | **38.56%** | **35.35%** | **56.44%** |
| | 29.47 | 6.05% | 93.95% | 28.50% | 38.48% | 33.01% | 58.12% |
| com-DBLP | **10** | **14.91%** | **85.09%** | **31.13%** | **41.65%** | **27.21%** | **79.14%** |
| | 10.01 | 14.91% | 85.09% | 31.13% | 41.65% | 27.21% | 79.14% |
| web-Stanford | **38** | **4.48%** | **95.52%** | **14.64%** | **58.23%** | **27.13%** | **77.87%** |
| | 37.74 | 5.05% | 94.95% | 16.16% | 58.19% | 25.61% | 78.04% |
| com-Youtube | **39** | **1.60%** | **98.40%** | **16.44%** | **50.92%** | **32.64%** | **55.93%** |
| | 36.09 | 1.75% | 98.25% | 17.58% | 50.87% | 31.55% | 56.75% |

## 5 Unknown network sampling SLSR

### 5.1 Traversal-based sampling algorithm at low sampling rates

Traversing on an original network establishes topological connections between sampled nodes, but may be overly attracted to the high-degree core nodes at low sampling rates. Thus, this section designs a new traversal-based sampling SLSR, which only adopts the information of the node set and the neighbors of sampled nodes, that is, any complex topological information of the unknown original network, such as, community, clique, and real statistical characteristics, cannot be used in design process of SLSR. To improve time efficiency, a low sampling rate $R_{SLSR}$ is needed, but the sampled subgraph should capture more properties of the original network.

Our SLSR creates a core-periphery framework for existing traversal-based samplings that run on unknown networks, such as FF [19], SRW [11], NBRW [20], CNRW [20], CNARW [20], and RD [21]. First, choose one sampling from the existing methods and use $T_s$ to represent it, then SLSR restricts $T_s$ to only traverse on the periphery using the evaluated $\ddot{d}_{org}$. Specifically, the set of neighbors of a node $v$ that is accessing by $T_s$ is changed as

$$N_{org}^{per}(v) = \{u | u \in N_{org}(v) \wedge d_{org}(u) \leq \ddot{d}_{org}\} \quad (2)$$

and other principles and steps of $T_s$ remain unchanged. Please note that, the other neighbor set

$$N_{org}^{cor}(v) = \{u | u \in N_{org}(v) \wedge d_{org}(u) > \ddot{d}_{org}\} \quad (3)$$

that is obtained simultaneously with Eq. (2) should be saved for the core sampling. The process of $T_s$ running on the periphery of $G_{org}$ with a sampling rate $R_{SLSR}$ is represented as follows:

$$G_{sub}^{per} = (V_{sub}^{per}, E_{sub}^{per}), \{N_{org}^{cor}(v) | v \in V_{sub}^{per}\} \leftarrow PeripherySampling(T_s, G_{org}, \ddot{d}_{org}, R_{SLSR}) \quad (4)$$

where $G_{sub}^{per}$ denotes the sampled subgraph of the periphery, and $V_{sub}^{per}$ and $E_{sub}^{per}$ denote its node set and edge set. According to P3, predefine a parameter $x\%$ to preserve top-$(\|N_{org}^{cor}(v)\| \times x\%)$ highest degree core neighbors for each $v \in V_{sub}^{per}$, and let $V_x(v)$ denote the node set composed of the preserved core neighbors where $V_x(v) \subseteq N_{org}^{cor}(v)$.

$$V_{sub}^{cor} = \bigcup_{v \in V_{sub}^{per}} V_x(v), \quad (5)$$

$$E_{sub}^{ver} = \{(v,u) | v \in V_{sub}^{per} \wedge u \in V_x(v)\}. \quad (6)$$

We define $V_{sub}^{cor}$ in Eq. (5) as the set of sampled core nodes. According to P2, all the connections between the $V_{sub}^{cor}$ nodes in $G_{org}$ should be preserved, which can be implemented by the access of the neighbors of each $V_{sub}^{cor}$ node in $G_{org}$. Please note that the $V_{sub}^{cor}$ nodes and the connections between them construct the sampled core. To ensure the connectivity of the sampled subgraph, the vertical edge set $E_{sub}^{ver}$ defined in Eq. (6) also should be preserved.

---

**Bisection method: Determining the parameter $x\%$**

1: **Input:** The sampled periphery $G_{sub}^{per} = (V_{sub}^{per}, E_{sub}^{per})$, the core neighbor sets $\{N_{org}^{cor}(v) | v \in V_{sub}^{per}\}$, the original network $G_{org} = (V_{org}, E_{org})$, and $\bar{d}_{org}$ that is the evaluated AD of $G_{org}$.

2: **Output:** $x\%$.

3: Initialize $x_{min} \leftarrow 1$, $x_{max} \leftarrow 100$, and $min\_distance \leftarrow 1000$.

4: **While** $x_{max} - x_{min} > 1$ **do**

5:   Update $x \leftarrow [(x_{max} + x_{min})/2]$, and $x\% \leftarrow x/100$, where $[X]$ rounds the element $X$ to the nearest integer.

6:   Initialize $V_{sub}^{cor} \leftarrow \emptyset$ and $E_{sub}^{ver} \leftarrow \emptyset$.

7:   **For** each node $v \in V_{sub}^{per}$, assume $N_{org}^{cor}(v) = \{u_1, u_2, \cdots, u_t\}$ with $d_{org}(u_1) \geq d_{org}(u_2) \geq \cdots \geq d_{org}(u_t)$, where $t = \|N_{org}^{cor}(v)\|$.

8:     Derive $k \leftarrow [\|N_{org}^{cor}(v)\| \times x\%]$, and update $V_{sub}^{cor} \leftarrow V_{sub}^{cor} \cup \{u_1, u_2, \cdots, u_k\}$ and $E_{sub}^{ver} \leftarrow E_{sub}^{ver} \cup \{(v, u_1), (v, u_2), \cdots, (v, u_k)\}$.

9:   **End For**

10:  Derive a sampled core $G_{sub}^{cor} = (V_{sub}^{cor}, E_{sub}^{cor})$ that preserves all the connections between $V_{sub}^{cor}$ nodes in $G_{org}$, and a sampled subgraph $G_{sub} = (V_{sub}, E_{sub})$ where $V_{sub} = V_{sub}^{cor} \cup V_{sub}^{per}$ and $E_{sub} = E_{sub}^{cor} \cup E_{sub}^{per} \cup E_{sub}^{ver}$.

11:  Derive the average degree $\bar{d}_{sub} = 2\|E_{sub}\|/\|V_{sub}\|$ of $G_{sub}$.

12:  **If** $\bar{d}_{sub} \geq \bar{d}_{org}$, **then** update $x_{max} \leftarrow x$, **Else** update $x_{min} \leftarrow x$. **End If**

13:  **If** $|\bar{d}_{sub} - \bar{d}_{org}| < min\_distance$, **then** update $x_{opt} \leftarrow x$ and $min\_distance \leftarrow |\bar{d}_{sub} - \bar{d}_{org}|$. **End If**

14: **End While**

15: Output $x\%$ where $x = x_{opt}$.

Next, analyze how to determine the parameter $x\%$. With the growth of $x\%$, the number of the vertical edges (namely, $\|E_{sub}^{ver}\|$) increases, while the sampled periphery $G_{sub}^{per}$ remains unchanged and the sampled core induced by the node set $V_{sub}^{cor}$ has a low variability [15]. Thus, the AD of the subgraph, composed of the vertical edges, the sampled periphery, and the sampled core, increases monotonically as $x\%$ grows, that is, a bisection method can be used to determine $x\%$ under the guidance of the evaluated $\bar{d}_{org}$ that is our target AD of the sampled subgraph.

In the bisection method proposed on the previous page, the number of iterations of the While loop is not more than $\log_2 100$. In addition, $\|N_{org}^{cor}(v)\| \leq \ddot{d}_{org}$ for each $v \in V_{sub}^{per}$ in line 7, and $\|V_{sub}^{cor}\| \ll \|V_{sub}^{per}\|$ in line 10, since the scale of the core is much smaller than that of the periphery, as shown in Table 4. Thus, the time complexity of the bisection method is $O(\|V_{sub}^{per}\|)$. Once the parameter $x\%$ is determined, the sampled subgraph of $G_{org}$ can be obtained using lines 6 to 10 in the bisection method. The Main function of SLSR is described as follows:

---

**SLSR: Main function**

---

1: **Input:** A sampling rate $R_{SLSR}$, and an original network $G_{org} = (V_{org}, E_{org})$ with unknown topological information but the node set and the neighbors of sampled nodes in the network can be accessed.

2: **Output:** $G_{sub} = (V_{sub}, E_{sub})$, namely a sampled subgraph of $G_{org}$, where $V_{sub}$ and $E_{sub}$ respectively denote the sampled node set and edge set.

3: Evaluate the AD value $\bar{d}_{org}$ and the DT value $\ddot{d}_{org}$ of $G_{org}$ using the random node sampling designed in Section 4.1 with $R_{RN} = 35\%$.

4: Choose an existing traversal-based sampling $T_s$, and run $T_s$ on the periphery of $G_{org}$ based on $\ddot{d}_{org}$:

$$G_{sub}^{per} = (V_{sub}^{per}, E_{sub}^{per}), \{N_{org}^{cor}(v)|v \in V_{sub}^{per}\} \leftarrow PeripherySampling(T_s, G_{org}, \ddot{d}_{org}, R_{SLSR}).$$

5: Determine the parameter $x\%$ using the bisection method proposed on the previous page.

6: Obtain $G_{sub} = (V_{sub}, E_{sub})$ under the constraint of $x\%$, and output the sampled subgraph.

Please note that the obtainment process is the same as lines 6 to 10 in the above-mentioned bisection method.

---

The sampling framework created by SLSR is simple. Simpler methods typically have higher time efficiency. However, the framework can significantly improve the accuracy of multi-structure preservation for $T_s$ at low sampling rates that are needed for a high time efficiency.

---

**PeripherySampling: Choosing $T_s$ as the FF sampling [19]**

---

1: **Input:** $G_{org} = (V_{org}, E_{org})$, $\ddot{d}_{org}$, and $R_{SLSR}$.

2: **Output:** $G_{sub}^{per} = (V_{sub}^{per}, E_{sub}^{per})$, and $\{N_{org}^{cor}(v)|v \in V_{sub}^{per}\}$.

3: Create an empty FIFO (first-in-first-out) queue $Q$ where $Q \leftarrow v$ represents adding $v$ to $Q$ and $v \leftarrow Q$ represents extracting and deleting $v$ from $Q$.

4: Derive $n = \|V_{org}\| \times R_{SLSR}$ that represents the expected number of periphery nodes to be sampled, and initialize $V_{sub}^{per} \leftarrow \emptyset$ and $E_{sub}^{per} \leftarrow \emptyset$.

5: Randomly choose a seed $w$ that falls in $\{v|v \in V_{org} \wedge d_{org}(v) \leq \ddot{d}_{org}\} - V_{sub}^{per}$, and $Q \leftarrow w$.

6: **While** $\|V_{sub}^{per}\| < n$ **do**:

7:     **If** $Q$ is empty, **then** randomly choose a seed $w$ that falls in $\{v|v \in V_{org} \wedge d_{org}(v) \leq \ddot{d}_{org}\} - V_{sub}^{per}$, and $Q \leftarrow w$. **End If**

8:     $v \leftarrow Q$; Derive $N_{org}^{cor}(v) = \{u|u \in N_{org}(v) \wedge d_{org}(u) > \ddot{d}_{org}\}$ and $N_{org}^{per}(v) = \{u|u \in N_{org}(v) \wedge d_{org}(u) \leq \ddot{d}_{org}\}$.

9:     Update $E_{sub}^{per} \leftarrow E_{sub}^{per} \cup \{(v,u)|u \in N_{org}^{per}(v) \cap V_{sub}^{per}\}$, and update $V_{sub}^{per} \leftarrow V_{sub}^{per} \cup \{v\}$,

10:     **If** $\|V_{sub}^{per}\| + \|Q\| < n$, **then** extract a fraction of nodes in $N_{org}^{per}(v) - V_{sub}^{per} - Q$,

    where the fraction is randomly drawn from a geometric distribution, and add the nodes into $Q$. **End If**

11: **End While**   %The probability of that $w$ in lines 5 and 7 falls into the periphery $\{v|v \in V_{org} \wedge d_{org}(v) \leq \ddot{d}_{org}\}$ is very high based on Table 4.

12: Output $G_{sub}^{per} = (V_{sub}^{per}, E_{sub}^{per})$ and $\{N_{org}^{cor}(v)|v \in V_{sub}^{per}\}$.

---

We arbitrarily choose $T_s$ as the FF sampling [19], since the sampling traverses each node no more than once which leads to its high time efficiency. Users can use $T_s$ to represent other existing traversal-based samplings. Once $T_s$ is chosen, the periphery sampling can be determined.

### 5.2 Analysis of the variance of sampled SLSR subgraphs

Low sampling rates may lead to high variances in the sampling results of large sample spaces. However, our SLSR sampling can control the variances based on the following three points:

- The bisection method is deterministic, that is, there is no randomness in the method.
- The random node sampling has a very high time efficiency, as shown in Table 3, thus setting its sampling rate $R_{RN} = 35\%$ can not only reduce the variance, but also has a weak impact on the time efficiency of our traversal-based SLSR sampling.
- The smaller the scale of a sample space, the lower the uncertainty of randomly extracting a sample from the space, which can be ensured by the theory of information entropy [16]. In the periphery sampling $T_s$ of SLSR, described in Section 5.1, the set of neighbors of a node $v$ that is accessing by $T_s$ has been compressed from $N_{org}(v)$ to $N_{org}^{per}(v) = \{u | u \in N_{org}(v) \wedge d_{org}(u) \leq \ddot{d}_{org}\}$. In a scale-free network, the scale of the neighbor set of a core node is extremely larger than that of a periphery node, as shown in Table 2, and the number of the vertical edges connecting a periphery node to a core node is much larger than the number of edges connecting a periphery node to another periphery node, as listed in Table 4. Thus, $\|N_{org}^{per}(v)\| < \|N_{org}(v)\| \ll \|N_{org}(u)\|$ when $v$ is a periphery node and $u$ is a core node. Please note that $N_{org}^{per}(v)$ is related to the sample space of $w$ when $T_s$ traverses from $v$ to $w \in N_{org}^{per}(v)$. Compared to the traditional traversal-based samplings that excessively prefer high-degree core nodes at low sampling rates, the sample space of $w$ in our SLSR sampling has been sharply compressed at most cases.

The above three points are critical for controlling the uncertainty of SLSR with $R_{SLSR} \leq 10\%$, and the experimental results in Section 6 can further verify the low variance.

## 6 Evaluation

### 6.1 Metrics

This paper proposes a traversal-based sampling SLSR that only uses the information of node set and the neighbors of sampled nodes, without involving complex topological characteristics, but can solve the issue of excessive preference for high-degree core nodes at low sampling rates. Thus, some metrics that can measure the excessive preference are included in this section.

**AD** is defined as $\sum k P(k) = 2\|E\|/\|V\|$ in a simple and undirected graph $G = (V, E)$ with node set $V$ and edge set $E$, where $P(k)$ denotes the fraction of nodes with degree $k$ in $G$ [29-31]. The statistic reflects whether a sampling favors core nodes with high-degrees.

**Complementry cumulative distribution** defined as $F(k)$ vs. $k$ where $F(k) = \sum_{d>k} P(d)$ exhibits better degree power-law characteristic [29].

**Average clustering coefficient (ACC)** defined as $\bar{C} = \sum C(k)P(k)$ represents how close a node's neighbors are to forming a clique [32-35], where $C(k) = 2T(k)/k(k-1)$ and $T(k)$ denotes the average of the number of links between two neighbors of $k$-degree nodes. A related distribution characteristic is **clustering coefficient distribution** [32] defined as $C(k)$ vs. $k$.

**Average path length (APL)** is defined as $\bar{L} = \sum l \cdot \mu(l)$ that represents the reachability of nodes within each other, where $\mu(l)$ denotes the fraction of node pairs with shortest path length $l$ between the two nodes. A related distribution characteristic is shortest **path length distribution** that is defined as $\mu(l)$ vs. $l$ [32,33,34].

**Ratio of weighted spectral distribution to node number (RWSD)** represents the connection relationship between low-degree nodes [36] most of which are in the periphery. The weighted spectral distribution is defined as $\sum(1-\lambda_i)^4$ where $\lambda_i$ denotes the $i^{th}$ eigenvalue in the normalized Laplacian spectrum of $G$ [36-38]. Please note that the statistic can be quickly calculated by a 4-cycle enumeration algorithm without the need for the calculation of the eigenvalues [39].

**Ratio of maximum degree to node number (RMD)** represents the influence of the node with maximum degree, which is suitable for the comparison of graphs with different scales [40].

**Closeness centrality (CC)** of a node is defined as $(n-1)/L$ where $n$ is the total number of nodes and $L$ denotes the sum of the length of the shortest path from the node to other nodes, which reflects how efficiently the node exchanges information with others [41].

**Betweenness centrality (BC)** of a node represents the fraction of the shortest paths that pass through the node for any pair of nodes, which describes potential power of the node in controlling the information flow in a network [42,43].

**Community** represents local densely-connected structures that are visually salient [44,45]. Thus, a visual evaluation was adopted. Specifically, the communities were detected by a Louvain method [45] and visually displayed by a force-directed method [46]. The correspondence between the communities of an original network and its sampled subgraphs was established by their shared nodes.

### 6.2 Original networks and sampled node number

The above multiple metrics and five widely-used large original networks chosen from Stanford Large Network Dataset Collection [28] were adopted for the evaluation. Please note that the original networks listed in Table 5 were simplified as undirected graphs, namely the self-loops, multi-edges, direction of edges, and a few isolated nodes with degree zero in the five chosen networks have been removed. In addition, the important statistics of the original networks have been listed in Tables 6 to 10 for the convenience of comparison.

**Table 5**. Descriptions of the five widely-used large original networks [28], and the mean and standard errors (below the mean) of the sampled node number of the SLSR subgraphs from 100 independent realizations for each sampling rate $R_{SLSR}$.

| Original networks | Original node number | Original edge number | Descriptions | $R_{SLSR}$ | Sampled node number |
|---|---|---|---|---|---|
| ego-Twitter | 81,305 | 1,342,220 | This dataset consists of 'circles' (or 'lists') from Twitter. | 5% | 5,733 ±172 |
| loc-Gowalla | 196,588 | 949,712 | This dataset consists of a location-based social network where users share their locations. | 5% | 11,256 ±159 |
| com-DBLP | 317,076 | 1,049,760 | This dataset consists of a co-authorship network from DBLP that is a computer science bibliography. | 10% | 49,402 ±155 |
| web-Stanford | 281,902 | 1,992,634 | This dataset consists of a web network from Stanford University (stanford.edu). | 10% | 30,385 ±294 |
| com-Youtube | 1,134,879 | 2,987,595 | This dataset consists of a social network from Youtube that is a video-sharing. | 5% | 58,998 ±232 |

According to Section 5.1, our SLSR sampling sequentially executes the AD and DT evaluation, the periphery sampling with a low sampling rate $R_{SLSR}$, and the bisection method for preserving core and vertical edges. Since the original network $G_{org} = (V_{org}, E_{org})$ is unknown, SLSR cannot know in advance the actual number of periphery nodes in $G_{org}$. Owing to that the number of core nodes is extremely smaller than that of periphery nodes, $\|V_{org}\| \times R_{SLSR}$ is approximately equal to the expected number of periphery nodes to be sampled. Thus, $R_{SLSR}$ is not a strict sampling rate that is defined as the ratio of the number of nodes in a sampled subgraph to $\|V_{org}\|$. In order to compare more fairly with the traditional traversal-based samplings on unknown networks, we first obtain the sampled SLSR subgraphs, and calculate the mean and standard errors of the sampled node number of the subgraphs from 100 independent realizations for each original network and given sampling rate $R_{SLSR}$ in Table 5. Then, each traditional traversal-based sampling outputs 100 subgraphs whose node number is equal to the mean of the sampled node number for each original network.

### 6.3 Comparisons

SLSR is compared with the related traversal-based samplings, namely FF [19], SRW [11], NBRW [20], CNRW [20], CNARW [20], and RD [21], on the five large original networks in Table 5, because the chosen samplings do not adopt complex topological characteristics, such as, community, clique, and real global statistical characteristics, of the original networks.

### 6.3.1 Variance comparison with statistics

At a low sampling rate, low variance is important for the reliability of sampling results. Thus, we first compare the standard errors of the statistics AD, ACC, AC, RWSD, RMD, and CC($v_{org}^{max}$) of the sampled subgraphs from 100 independent realizations, where $v_{org}^{max}$ is the maximum degree node in an original network and is easily preserved in the sampled subgraphs by the chosen sampling methods. Please note that, owing to the high time complexity, we compare the standard errors of the statistics APL and BC($v_{org}^{max}$) from 5 or 10 independent realizations.

**Table 6**. The statistics of the **original ego-Twitter network**, and the mean and standard errors (below the mean) of the statistics of the sampled subgraphs. APL and BC($v_{org}^{max}$) are related to 10 independent realizations, while other statistics are related to 100 independent realizations. The sampling rate $R_{SLSR}$ of our SLSR sampling, and the sampled node number of the other chosen samplings FF, RD, SRW, NBRW, CNRW, and CNARW have been illustrated in Section 6.2.

| Statistics | | AD | ACC | APL | RWSD | RMD | CC($v_{org}^{max}$) | BC($v_{org}^{max}$) |
|---|---|---|---|---|---|---|---|---|
| **Original network** | | **33.01** | **0.565** | **3.889** | **0.008** | **0.042** | **0.402** | **0.059** |
| Mean and standard errors of the sampled subgraphs | SLSR | 33.06 ±0.410 | 0.445 ±0.014 | 3.326 ±0.096 | 0.011 ±0.0008 | 0.137 ±0.035 | 0.442 ±0.017 | 0.056 ±0.012 |
| | FF | 46.26 ±6.895 | 0.512 ±0.018 | 3.058 ±0.149 | 0.007 ±0.0016 | 0.189 ±0.087 | 0.461 ±0.023 | 0.043 ±0.021 |
| | RD | 13.73 ±0.383 | 0.312 ±0.007 | 4.063 ±0.023 | 0.022 ±0.0009 | 0.088 ±0.007 | 0.387 ±0.003 | 0.101 ±0.006 |
| | SRW | 35.26 ±2.070 | 0.543 ±0.006 | 3.438 ±0.041 | 0.013 ±0.0009 | 0.111 ±0.011 | 0.440 ±0.007 | 0.060 ±0.007 |
| | NBRW | 34.856 ±1.918 | 0.544 ±0.006 | 3.428 ±0.078 | 0.013 ±0.0008 | 0.110 ±0.011 | 0.440 ±0.007 | 0.059 ±0.007 |
| | CNRW | 49.026 ±0.749 | 0.546 ±0.003 | 3.447 ±0.022 | 0.007 ±0.0002 | 0.081 ±0.003 | 0.443 ±0.003 | 0.055 ±0.003 |
| | CNARW | 37.65 ±1.651 | 0.467 ±0.005 | 3.306 ±0.032 | 0.012 ±0.0008 | 0.126 ±0.008 | 0.451 ±0.006 | 0.060 ±0.008 |

BC($v_{org}^{max}$) of the original com-Youtube network with 1,134,879 nodes and 2,987,595 edges was not provided in Table 10 due to extremely high computation and memory requirements. Please note that the RWSD of the network can be quickly obtained within 2 hours [39], whereas the APL and the path length distribution of the network have to be computed by a parallel algorithm. On a computer with Intel Core i7-8700 CPU 3.20 GHz Memory 16 G, the parallel algorithm with 5 threads used for calculating the APL and the path length distribution runs about 12 days.

Because $T_s$ in SLSR was chosen as the FF sampling for the experiments. Thus, we first compare the variance between SLSR and FF. Based on Tables 6 to 10, we can observe that the standard errors of the statistics of SLSR are generally smaller than those of FF, except for a few cases where the mean is very small. Please note that the low variance of SLSR has been analyzed in Section 5.2. NBRW, CNRW and CNARW are improved random walk samplings objective to reduce the asymptotic variance [10,20]. Although the three samplings provided rigorous mathematical proofs based on Markov chain, their theoretical basis is that the Markov chain must converge, which is difficult to be guaranteed at low sampling rates. Thus, based on the analysis in Section 5.2 and Tables 6 to 10, the standard errors of the statistics of SLSR can be effectively controlled in most cases. RD consists of two steps: the first step is to extract a predetermined number of starting seeds using the random node sampling, and the second step adopts a deterministic algorithm without randomness [21]. Thus, RD can also effectively control the standard errors in most cases.

Mean is another important indicator of the statistics. Thus, Section 6.3.2 will use the mean to analyze the excessive preference for high-degree core nodes at low sampling rates.

**Table 7.** The statistics of the **original loc-Gowalla network**, and the mean and standard errors (below the mean) of the statistics of the sampled subgraphs. APL and $BC(v_{org}^{max})$ are related to 10 independent realizations, while other statistics are related to 100 independent realizations. The sampling rate $R_{SLSR}$ of our SLSR sampling, and the sampled node number of the other chosen samplings FF, RD, SRW, NBRW, CNRW, and CNARW have been illustrated in Section 6.2.

| Statistics | | AD | ACC | APL | RWSD | RMD | $CC(v_{org}^{max})$ | $BC(v_{org}^{max})$ |
|---|---|---|---|---|---|---|---|---|
| **Original network** | | **9.66** | **0.236** | **4.627** | **0.057** | **0.075** | **0.389** | **0.324** |
| **Mean and standard errors of the sampled subgraphs** | SLSR | 9.700 ±0.194 | 0.236 ±0.007 | 4.644 ±0.242 | 0.071 ±0.011 | 0.096 ±0.031 | 0.376 ±0.024 | 0.312 ±0.091 |
| | FF | 24.12 ±4.044 | 0.280 ±0.008 | 3.483 ±0.232 | 0.021 ±0.009 | 0.205 ±0.051 | 0.490 ±0.041 | 0.253 ±0.027 |
| | RD | 11.48 ±0.105 | 0.219 ±0.003 | 3.505 ±0.016 | 0.051 ±0.002 | 0.288 ±0.002 | 0.497 ±0.002 | 0.375 ±0.003 |
| | SRW | 21.23 ±0.421 | 0.309 ±0.004 | 3.708 ±0.036 | 0.035 ±0.002 | 0.205 ±0.005 | 0.473 ±0.005 | 0.289 ±0.013 |
| | NBRW | 20.42 ±0.484 | 0.315 ±0.004 | 3.717 ±0.031 | 0.040 ±0.001 | 0.201 ±0.006 | 0.468 ±0.005 | 0.290 ±0.011 |
| | CNRW | 21.11 ±0.126 | 0.292 ±0.002 | 3.781 ±0.016 | 0.028 ±0.001 | 0.164 ±0.002 | 0.465 ±0.002 | 0.306 ±0.002 |
| | CNARW | 21.16 ±0.458 | 0.253 ±0.004 | 3.661 ±0.053 | 0.037 ±0.002 | 0.209 ±0.005 | 0.474 ±0.006 | 0.288 ±0.008 |

**Table 8.** The statistics of the **original com-DBLP network**, and the mean and standard errors (below the mean) of the statistics of the sampled subgraphs. APL and $BC(v_{org}^{max})$ are related to 5 independent realizations, while other statistics are related to 100 independent realizations. The sampling rate $R_{SLSR}$ of our SLSR sampling, and the sampled node number of the other chosen samplings FF, RD, SRW, NBRW, CNRW, and CNARW have been illustrated in Section 6.2.

| Statistics | | AD | ACC | APL | RWSD | RMD | $CC(v_{org}^{max})$ | $BC(v_{org}^{max})$ |
|---|---|---|---|---|---|---|---|---|
| **Original network** | | **6.621** | **0.632** | **6.792** | **0.070** | **0.001** | **0.218** | **0.007** |
| **Mean and standard errors of the sampled subgraphs** | SLSR | 6.231 ±0.052 | 0.557 ±0.002 | 6.852 ±0.013 | 0.098 ±0.001 | 0.002 ±0.0001 | 0.207 ±0.001 | 0.003 ±0.0003 |
| | FF | 9.887 ±0.134 | 0.564 ±0.005 | 5.576 ±0.047 | 0.048 ±0.001 | 0.005 ±0.0004 | 0.240 ±0.002 | 0.002 ±0.0007 |
| | RD | 8.013 ±0.023 | 0.313 ±0.002 | 5.688 ±0.009 | 0.050 ±0.001 | 0.005 ±0.0001 | 0.237 ±0.0003 | 0.002 ±0.0001 |
| | SRW | 8.975 ±0.068 | 0.592 ±0.003 | 6.075 ±0.017 | 0.061 ±0.001 | 0.005 ±0.0002 | 0.239 ±0.001 | 0.012 ±0.0012 |
| | NBRW | 8.769 ±0.076 | 0.616 ±0.002 | 6.156 ±0.021 | 0.065 ±0.001 | 0.005 ±0.0002 | 0.236 ±0.001 | 0.012 ±0.0009 |
| | CNRW | 9.001 ±0.067 | 0.591 ±0.003 | 6.055 ±0.032 | 0.060 ±0.001 | 0.005 ±0.0002 | 0.239 ±0.001 | 0.012 ±0.0009 |
| | CNARW | 7.928 ±0.055 | 0.467 ±0.002 | 5.936 ±0.006 | 0.062 ±0.001 | 0.004 ±0.0002 | 0.241 ±0.001 | 0.009 ±0.001 |

Table 9. The statistics of the **original web-Stanford network**, and the mean and standard errors (below the mean) of the statistics of the sampled subgraphs. APL and BC($v_{org}^{max}$) are related to 10 independent realizations, while other statistics are related to 100 independent realizations. The sampling rate $R_{SLSR}$ of our SLSR sampling, and the sampled node number of the other chosen samplings FF, RD, SRW, NBRW, CNRW, and CNARW have been illustrated in Section 6.2.

| Statistics | | AD | ACC | APL | RWSD | RMD | CC($v_{org}^{max}$) | BC($v_{org}^{max}$) |
|---|---|---|---|---|---|---|---|---|
| **Original network** | | **14.14** | **0.597** | **6.815** | **0.049** | **0.133** | **0.279** | **0.632** |
| **Mean and standard errors of the sampled subgraphs** | SLSR | 14.22 ±0.404 | 0.568 ±0.011 | 5.246 ±0.127 | 0.038 ±0.002 | 0.173 ±0.005 | 0.355 ±0.010 | 0.684 ±0.022 |
| | FF | 26.80 ±2.069 | 0.566 ±0.025 | 4.025 ±0.268 | 0.024 ±0.003 | 0.297 ±0.058 | 0.465 ±0.034 | 0.663 ±0.060 |
| | RD | 5.629 ±0.312 | 0.321 ±0.020 | 4.987 ±0.116 | 0.058 ±0.002 | 0.169 ±0.013 | 0.368 ±0.013 | 0.704 ±0.024 |
| | SRW | 27.18 ±4.710 | 0.678 ±0.048 | 4.209 ±0.269 | 0.016 ±0.002 | 0.368 ±0.076 | 0.430 ±0.062 | 0.677 ±0.051 |
| | NBRW | 26.72 ±3.887 | 0.675 ±0.055 | 4.621 ±0.598 | 0.017 ±0.003 | 0.355 ±0.058 | 0.421 ±0.051 | 0.680 ±0.065 |
| | CNRW | 27.12 ±4.589 | 0.689 ±0.037 | 4.226 ±0.294 | 0.015 ±0.002 | 0.359 ±0.076 | 0.432 ±0.058 | 0.656 ±0.092 |
| | CNARW | 22.71 ±1.767 | 0.553 ±0.035 | 4.420 ±0.170 | 0.021 ±0.002 | 0.328 ±0.043 | 0.407 ±0.032 | 0.584 ±0.084 |

Table 10. The statistics of the **original com-Youtube network**, and the mean and standard errors (below the mean) of the statistics of the sampled subgraphs. APL and BC($v_{org}^{max}$) are related to 5 independent realizations, while other statistics are related to 100 independent realizations. The sampling rate $R_{SLSR}$ of our SLSR sampling, and the sampled node number of the other chosen samplings FF, RD, SRW, NBRW, CNRW, and CNARW have been illustrated in Section 6.2.

| Statistics | | AD | ACC | APL | RWSD | RMD | CC($v_{org}^{max}$) | BC($v_{org}^{max}$) |
|---|---|---|---|---|---|---|---|---|
| **Original network** | | **5.265** | **0.081** | **5.279** | **0.081** | **0.025** | **0.338** | — |
| **Mean and standard errors of the sampled subgraphs** | SLSR | 5.201 ±0.174 | 0.092 ±0.002 | 5.808 ±0.099 | 0.136 ±0.004 | 0.055 ±0.002 | 0.257 ±0.003 | — |
| | FF | 21.90 ±0.532 | 0.128 ±0.005 | 3.971 ±0.052 | 0.027 ±0.003 | 0.118 ±0.003 | 0.331 ±0.003 | — |
| | RD | 9.857 ±0.750 | 0.099 ±0.005 | 4.223 ±0.008 | 0.067 ±0.005 | 0.117 ±0.005 | 0.321 ±0.002 | — |
| | SRW | 19.25 ±2.378 | 0.145 ±0.015 | 4.041 ±0.014 | 0.037 ±0.007 | 0.127 ±0.009 | 0.432 ±0.019 | — |
| | NBRW | 19.29 ±0.987 | 0.143 ±0.003 | 4.133 ±0.009 | 0.047 ±0.004 | 0.122 ±0.002 | 0.424 ±0.012 | — |
| | CNRW | 19.97 ±0.138 | 0.141 ±0.002 | 4.020 ±0.012 | 0.036 ±0.001 | 0.125 ±0.002 | 0.435 ±0.002 | — |
| | CNARW | 19.11 ±2.296 | 0.112 ±0.010 | 4.215 ±0.482 | 0.040 ±0.007 | 0.121 ±0.015 | 0.428 ±0.018 | — |

### 6.3.2 Mean comparison with statistics

This section first analyzes the importance of the statistics AD, ACC, APL, RWSD, and RMD in measuring the excessive preference for high-degree core nodes, and then experimentally studies the influence of the core-periphery structures on the excessive preference.

Scale-free networks consist of a dense core and a sparse periphery, and the core is densely connected by the periphery. Thus, high-degree core nodes tend to be densely connected to each other. If more high-degree core nodes are sampled, the subgraph induced by the sampled nodes can preserve more edges, that is, AD becomes larger as the preference is stronger.

Given a periphery node $v$, owing to the dense core, the local clustering coefficient of $v$ tend to be larger as the proportion of high-degree core nodes in $N_{org}(v)$ increases. Thus, ACC generally becomes larger as the preference is stronger.

Owing to the sparse periphery, nodes in the periphery have to shorten the path length between each other through the core, that is, APL becomes smaller as the preference is stronger.

According to the study of Jiao et al. [36], RWSD indicates the feature of connections between low-degree nodes on large networks, and the statistic decreases as the connections become sparser. The excessive preference for core nodes can make the connections between low-degree periphery nodes much sparser, which lead to smaller RWSD.

In addition, the excessive preference for core nodes generally induces larger RMD because the maximum degree node must be located in the core.

According to the mean values of the above-mentioned statistics and the core-periphery structures of the original networks shown in Table 4, we study the influence of the core structures on the excessive preference for the core nodes, which is an important contribution of this paper, because it is helpful in improving the traditional traversal-based samplings at low sampling rates.

Table 4 shows that the com-Youtube network has the smallest core node percentage and largest core edge density, while the mean values of the statistics in Table 10 confirm that the traditional traversal-based samplings have the strongest excessive preference for the core. Conversely, Table 4 shows that the com-DBLP network owns the largest core node percentage and smallest core edge density, while the mean values of the statistics in Table 8 verify that the traditional traversal-based samplings can avoid the excessive preference for the core. According to Tables 6 to 10, our SLSR sampling performs outstandingly in Tables 7, 9 and 10. The results have been analyzed in Section 4.5, that is, the cores of the com-Youtube, web-Stanford and loc-Gowalla networks in Table 4 are much smaller and denser than those of other chosen networks.

Moreover, we compare two centralities (i.e., CC and BC) of some important nodes between the original networks and their sampled subgraphs. The nodes were chosen as the maximum degree nodes $v_{org}^{max}$ in the original networks, and they can be preserved to the sampled subgraphs by the biased samplings being compared. The chosen $T_s$ in SLSR is competent for capturing peripheral degree distribution without the interference of the core, all the edges that connect $v_{org}^{max}$ to nodes in $V_{sub}^{per}$ are preserved due to P3, where $G_{sub}^{per} = (V_{sub}^{per}, E_{sub}^{per})$ is the output of $T_s$, and the structure of the core is preserved to the maximum extent possible due to P2. All the clues strongly influence the centralities of the top highest degree core nodes in the SLSR sampled subgraphs, and Tables 6 to 10 verify that the clues are helpful in capturing $\mathbf{CC}(v_{org}^{max})$ and $\mathbf{BC}(v_{org}^{max})$.

### 6.3.3 Distribution comparison

This section chooses degree complementary cumulative distribution [29], clustering coefficient distribution [32], and path length distribution [32,34], which are commonly-used measures, for further comparison. Because it is difficult to display the distributions of 100 realizations simultaneous, we choose only one realization with AD closest to the mean for the comparison. In addition, we choose top five sampling methods with minimum AD standard errors for each original network, since high variance corresponds to high uncertainty.

The distribution of nodes with lower degrees is more important for the first two measures. For example, in Fig. 4(*a*) and (*b*), degrees not exceeding 10 correspond to 85.09% of the total number of nodes in the com-DBLP network. The path length distribution of a graph was calculated on the maximum connected component of the graph.

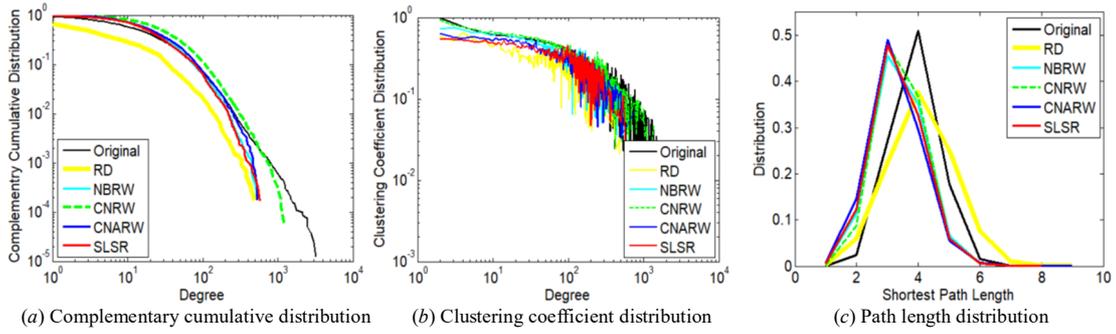

(*a*) Complementary cumulative distribution    (*b*) Clustering coefficient distribution    (*c*) Path length distribution

**Fig. 2.** Comparison of distribution characteristics between **original ego-Twitter network** and its subgraphs sampled by SLSR and related methods with relatively low variances. The sampled node number of the subgraphs has been illustrated in Section 6.2.

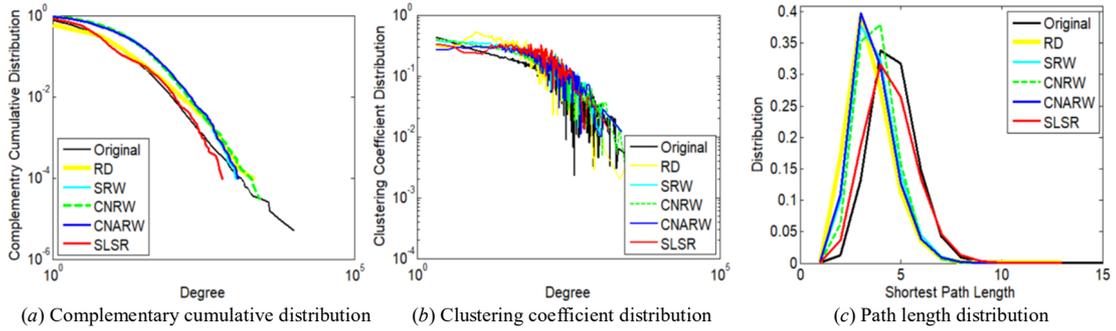

(*a*) Complementary cumulative distribution    (*b*) Clustering coefficient distribution    (*c*) Path length distribution

**Fig. 3.** Comparison of distribution characteristics between **original loc-Gowalla network** and its subgraphs sampled by SLSR and related methods with relatively low variances. The sampled node number of the subgraphs has been illustrated in Section 6.2.

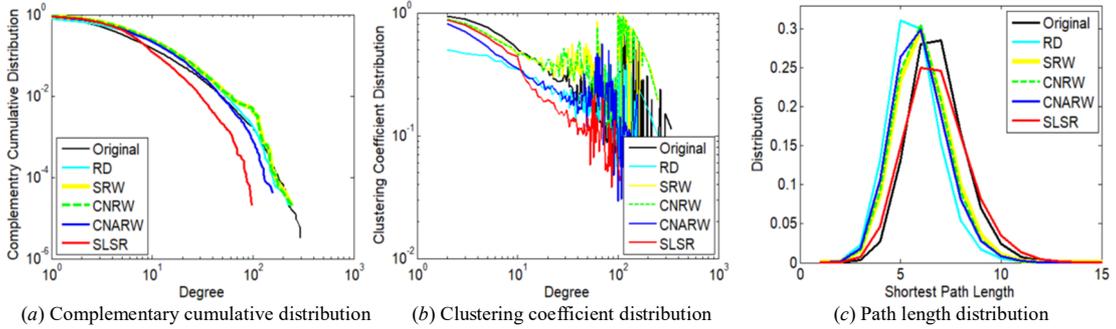

(*a*) Complementary cumulative distribution    (*b*) Clustering coefficient distribution    (*c*) Path length distribution

**Fig. 4.** Comparison of distribution characteristics between **original com-DBLP network** and its subgraphs sampled by SLSR and related methods with relatively low variances. The sampled node number of the subgraphs has been illustrated in Section 6.2.

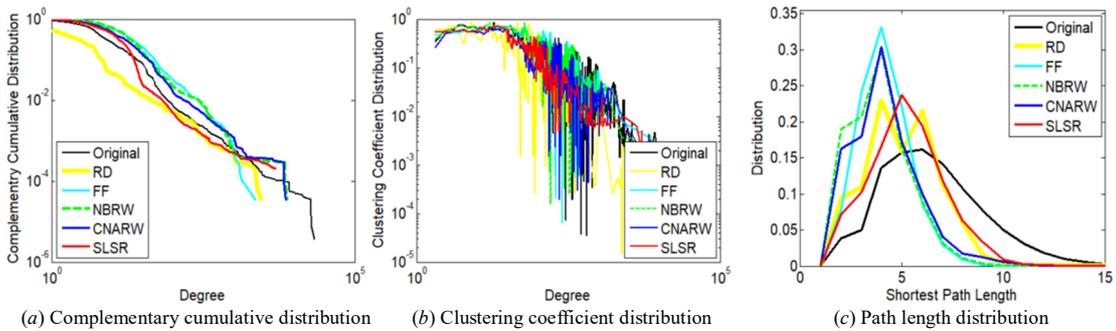

(*a*) Complementary cumulative distribution    (*b*) Clustering coefficient distribution    (*c*) Path length distribution

**Fig. 5.** Comparison of distribution characteristics between **original web-Stanford network** and its subgraphs sampled by SLSR and related methods with relatively low variances. The sampled node number of the subgraphs has been illustrated in Section 6.2.

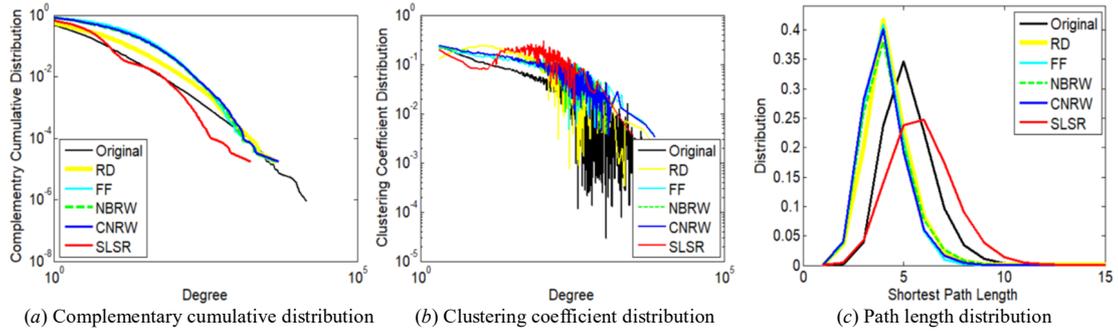

(a) Complementary cumulative distribution  (b) Clustering coefficient distribution  (c) Path length distribution

**Fig. 6.** Comparison of distribution characteristics between **original com-Youtube network** and its subgraphs sampled by SLSR and related methods with relatively low variances. The sampled node number of the subgraphs has been illustrated in Section 6.2.

Based on the comparisons of Figs. 2 to 6, SLSR can preserve the three distributions.

### 6.3.4 Community visualization

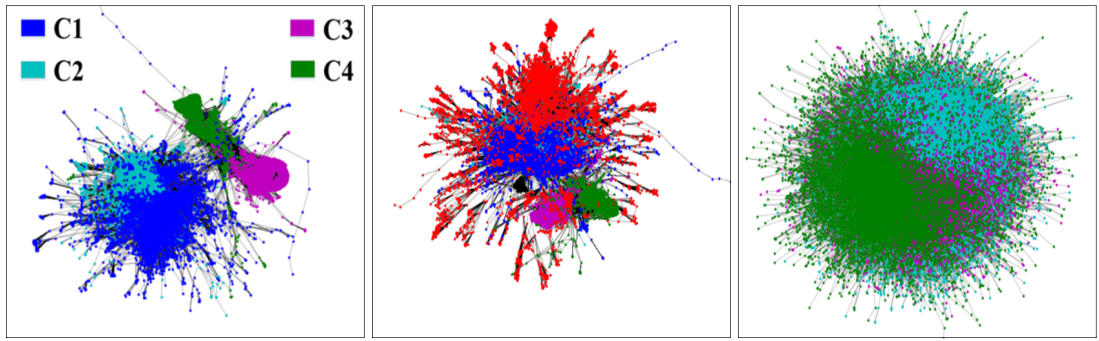

(a) Top 4 largest communities   (b) Top 6 largest communities   (c) Top 3 largest communities

**Fig. 7.** Visualization of top largest communities in the (a)(b) original web-Stanford network and the (c) original loc-Gowalla network. Specifically, the top 4 largest communities in (a) are named as C1, C2, C3 and C4, respectively.

Force-directed layout [46] is a powerful visualization tool for communities, but has two shortcomings when applied to large networks. One is that communities overlap severely with each other, the other is that the layout speed is extremely slow. To make up for the shortcomings, we first use a Louvain method [45] to detect the communities of an original network and its sampled subgraphs, and then extract top-$k$ largest communities and visualize them. As shown in Fig.7, the boundaries of distinct communities in the original web-Stanford network are clearer than those in the original loc-Gowalla network. Thus, we chose the former for the visualization comparison.

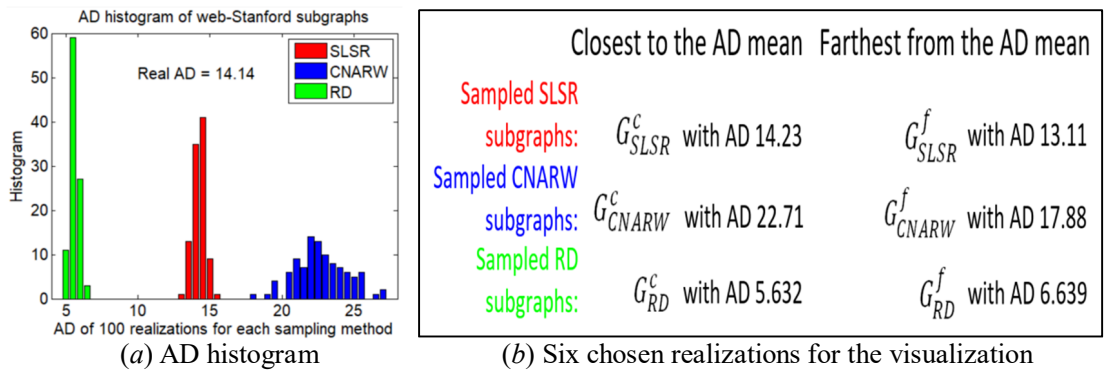

(a) AD histogram   (b) Six chosen realizations for the visualization

**Fig. 8.** AD histograms of 100 realizations for each method (i.e., SLSR, CNARW and RD) on the original web-Stanford network, and the chosen realizations for the visualization. The sampled node number of the three methods has been illustrated in Section 6.2.

The difficulty of the visualization under low sampling rates lies in the uncertainty of sampling results induced by high variances. To solve this issue, we only visualize the subgraphs obtained by RD, SLSR, and CNARW, which exhibit the lowest AD standard errors in Table 9. In addition, we choose two realizations for each sampling method, one with AD closest to the mean and the other with AD farthest from the mean, as shown in Fig. 8.

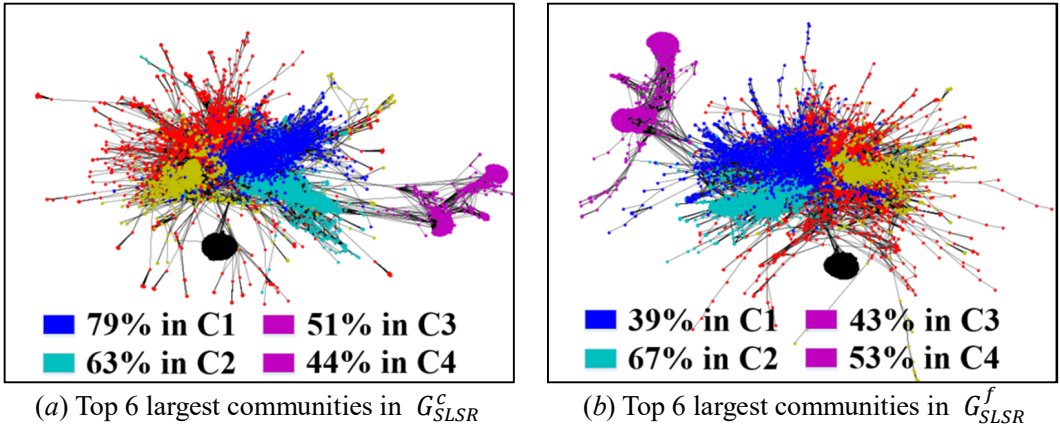

(a) Top 6 largest communities in $G_{SLSR}^{c}$    (b) Top 6 largest communities in $G_{SLSR}^{f}$

**Fig. 9.** Visualization of communities in the subgraphs $G_{SLSR}^{c}$ with AD 14.23 and $G_{SLSR}^{f}$ with AD 13.11 that were sampled by **SLSR** from the original web-Stanford network and were illustrated in Fig. 8. (a) Top 6 largest communities in $G_{SLSR}^{c}$, in which 79% of blue nodes fall in C1, 63% of cyan nodes fall in C2, 51% of magenta nodes fall in C3, and 44% of magenta nodes fall in C4. (b) Top 6 largest communities in $G_{SLSR}^{f}$, in which 39% of blue nodes fall in C1, 67% of cyan nodes fall in C2, 43% of magenta nodes fall in C3, and 53% of magenta nodes fall in C4.

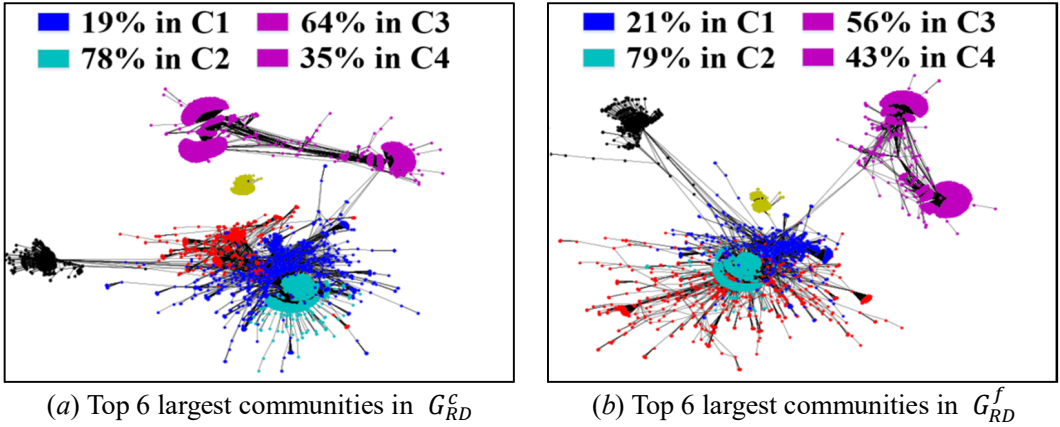

(a) Top 6 largest communities in $G_{RD}^{c}$    (b) Top 6 largest communities in $G_{RD}^{f}$

**Fig. 10.** Visualization of communities in the subgraphs $G_{RD}^{c}$ with AD 5.632 and $G_{RD}^{f}$ with AD 6.639 that were sampled by **RD** from the original web-Stanford network and were illustrated in Fig. 8. (a) Top 6 largest communities in $G_{RD}^{c}$, in which 19% of blue nodes fall in C1, 78% of cyan nodes fall in C2, 64% of magenta nodes fall in C3, and 35% of magenta nodes fall in C4. (b) Top 6 largest communities in $G_{RD}^{f}$, in which 21% of blue nodes fall in C1, 79% of cyan nodes fall in C2, 56% of magenta nodes fall in C3, and 43% of magenta nodes fall in C4.

Let $C_i = (V_{ci}, E_{ci})$ denote a community in Fig. 7(a), and $S = (V_s, E_s)$ denote a community in Figs. 9 to 11. If $r_S = \|V_s \cap V_{ci}\|/\|V_s\| \geq 19\%$, we believe that $S$ in the sampled subgraphs originates from $C_i$ in the original network, because Fig. 7(a) and (b) show that $C_1$ and $C_2$ gather with other communities where the boundaries between them are vague, and the Louvain method [45] used for community detection is a random algorithm and may divide the vague boundaries into different communities. In addition, Fig. 7(a) and (b) show that $C_3$ and $C_4$ are remote communities far from the gathering center that includes $C_1$ and $C_2$.

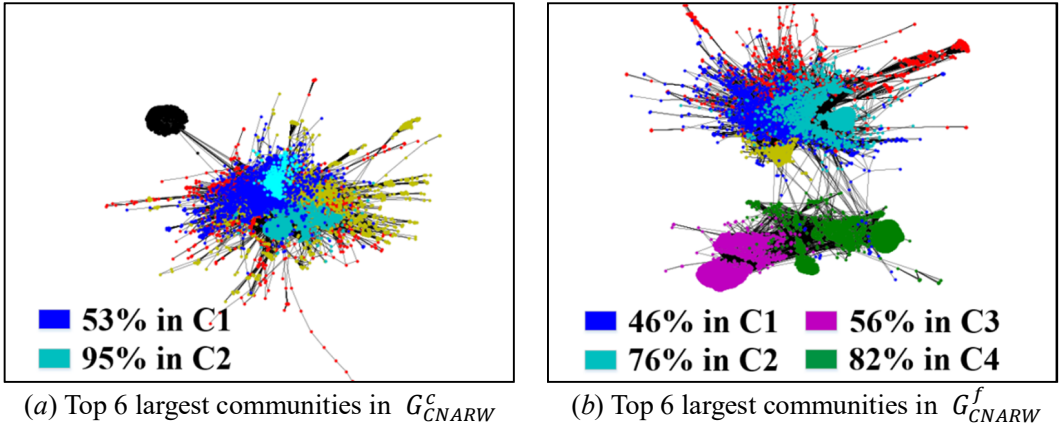

(a) Top 6 largest communities in $G^c_{CNARW}$  (b) Top 6 largest communities in $G^f_{CNARW}$

**Fig. 11.** Visualization of communities in the subgraphs $G^c_{CNARW}$ with AD 22.71 and $G^f_{CNARW}$ with AD 17.88 that were sampled by **CNARW** from the original web-Stanford network and were illustrated in Fig. 8. (*a*) Top 6 largest communities in $G^c_{CNARW}$, in which 53% of blue nodes fall in C1, 95% of cyan nodes fall in C2. (*b*) Top 6 largest communities in $G^f_{CNARW}$, in which 46% of blue nodes fall in C1, 76% of cyan nodes fall in C2, 56% of magenta nodes fall in C3, and 82% of green nodes fall in C4.

Both SLSR and RD adopt the random node sampling that randomly chooses nodes with uniform distribution, which can effectively avoid getting stuck locally at low sampling rates, as shown in Figs. 9 and 10. Different from RD, SLSR separates the random node sampling from the subgraph representation, that is, the random node sampling is only used for the evaluation of AD and DT, and does not participate in the obtainment of subgraphs. Specifically, SLSR uses the periphery sampling to construct the complex topological connections between periphery nodes, and adopts the bisection method to preserve important high-degree core nodes. Thus, SLSR not only inherits the advantage of the random node sampling in avoiding getting stuck locally, but also compensates for the shortcomings of the random node sampling analyzed in Sections 4.2 and 4.3, that is, the random node sampling not only loses top highest degree nodes in the core but also ignores the complex topological correlation between sampled periphery nodes.

The initial seed in Fig. 11(*a*) falls in the gathering center including $C_1$ and $C_2$, while the huge magenta and green communities in Fig. 11(b) are caused by the initial seed falling into $C_3$ or $C_4$ in Fig. 7(*a*). Although the random walk-based samplings [10,11,20], such as, SRW, NBRW, CNARW, and CNRW, focus on the theoretical proof of asymptotic variance, mathematical theory is usually based on simplified assumptions of the real world, and Figs. 8(*a*),11 and Table 10 experimentally show that the low sampling rates do not meet the simplified assumptions of the Markov chain theory.

### 6.3.5 Time efficiency of the traversal-based samplings

The seven sampling methods, namely, SLSR, FF, SRW, NBRW, CNRW, CNARW and RD, run on another computer with Intel Core i7-8550U CPU 1.80 GHz Memory 20 G. The time comparisons of the sampling methods are listed in Table 11, which shows that SLSR maintains high time efficiency of unknown graph samplings. The running time of SLSR depends on the bisection method and the periphery sampling designed in Section 5. Specifically, the periphery sampling corresponds to the chosen $T_s$ sampling, and the time complexity of the bisection method is restricted by $\|V^{per}_{sub}\| \times \log_2 100$ where $\|V^{per}_{sub}\|$ denotes the number of sampled periphery nodes that has been sharply decreased in contrast to the number of nodes in the original network $G_{org} = (V_{org}, E_{org})$ owing to the low sampling rate $R_{SLSR} \leq 10\%$. Please note that the $T_s$ sampling can be replaced by other sampling methods with high time efficiency in Table 11.

Table 11. The mean and standard errors (below the mean) of running time (Seconds) of SLSR and related methods from 100 independent realizations that sample the original networks in Table 5. The sampled node number has been illustrated in Section 6.2.

| Original networks | | ego-Twitter | loc-Gowalla | com-DBLP | web-Stanford | com-Youtube |
|---|---|---|---|---|---|---|
| Original node number | | 81,305 | 196,588 | 317,076 | 281,902 | 1,134,879 |
| Sampled node number | | 5,733 | 11,256 | 49,402 | 30,385 | 58,998 |
| | | **Runtime (Seconds)** | | | | |
| Sampling methods | SLSR | 7.079 ±0.649 | 16.06 ±1.061 | 325.5 ±41.65 | 77.11 ±8.605 | 190.5 ±9.888 |
| | FF | 8.126 ±0.583 | 23.58 ±2.268 | 241.6 ±23.26 | 114.6 ±9.659 | 309.5 ±29.03 |
| | RD | 2.776 ±0.313 | 6.325 ±0.505 | 9.322 ±0.768 | 24.13 ±3.515 | 33.69 ±3.233 |
| | SRW | 8.100 ±0.689 | 7.660 ±0.699 | 9.810 ±1.390 | 25.25 ±6.881 | 39.39 ±4.507 |
| | NBRW | 9.590 ±1.484 | 8.580 ±1.628 | 8.880 ±1.776 | 22.15 ±5.585 | 39.90 ±5.217 |
| | CNRW | 11.64 ±1.715 | 10.43 ±0.997 | 13.09 ±0.943 | 17.74 ±2.769 | 41.46 ±3.245 |
| | CNARW | 58.45 ±3.245 | 56.30 ±5.062 | 58.33 ±3.411 | 353.3 ±45.13 | 329.6 ±32.75 |

SLSR chose FF as $T_s$ in the experiments. Based on Table 11, SLSR can enhance the time efficiency of FF at most cases except for the case running on the com-DBLP network. According to the variance analysis in Section 5.2, $N_{org}^{per}(v)$ is the sample space of $w$ when $T_s$ traverses from $v$ to $w \in N_{org}^{per}(v)$, and $\|N_{org}^{per}(v)\| < \|N_{org}(v)\| \ll \|N_{org}(u)\|$ where $v$ is a periphery node and $u$ is a core node. The compressed sample space not only reduces the uncertainty, but also enhances the time efficiency; for example, extracting one node from 10 nodes is definitely faster than extracting from 1000 nodes. However, as shown in Table 4, the degree threshold $\ddot{d}_{org} = 10$ of the original com-DBLP network is much smaller than those of other original networks. Note that $d_{org}(v) \leq \ddot{d}_{org}$ for each node $v$ in the periphery, where $d_{org}(v) = \|N_{org}(v)\|$ is the degree of node $v$ in the original network $G_{org} = (V_{org}, E_{org})$; thus, the sample space of $w$ when $T_s$ traverses from $v$ to $w$ is already very small in the original com-DBLP network. In addition, Table 4 shows that the com-DBLP network owns the largest core node percentage and smallest core edge density in contrast to those of other networks, which is helpful in avoiding the excessive preference for the core, as shown in Table 8. Thus, the runtime of the periphery sampling of SLSR (excluding the bisection method) is close to the runtime of FF, when running on the original com-DBLP network.

## 7 Conclusions

Low sampling rates redefine the concept of "large networks" from thousands of nodes to millions of nodes or more nodes, which are indispensable for large network analysis. In addition, the concept of "unknown" prevents the sampling methods from using complex topological information, such as, community, clique, and real statistical characteristics of the original networks, which provides a guarantee for high time efficiency on large networks.

The contributions of this paper focus on the "low sampling rates" and "unknown", as well as the uncertainty and variance problem at the low sampling rates. Specifically,

- We analyze the advantages and shortcomings of the random node sampling, separate the random node sampling from the subgraph representation, and design a simple and deterministic bisection method after the random node evaluation and periphery sampling. The designed method not only inherits the advantage of the random node sampling in avoiding getting stuck locally, but also preserves the periphery topological structure, critical high-degree core nodes, and edges induced by the core nodes.
- We theoretically analyze the impact of the sample space of $w$ (when $T_s$ traverses from $v$ to $w$) on entropy (uncertainty) and time efficiency, and experimentally verify the low variance and high time efficiency of our traversal-based SLSR sampling.
- We investigate a blind spot in Markov chain theory, namely, the low sampling rates, and find that, the smaller and denser the cores in original networks, the stronger the preference of the traditional traversal-based samplings for high-degree core nodes. The new finding is helpful in addressing the issues caused by the low sampling rates.
- We adopt a local visualization method to experimentally verify the characteristics of low variance and community capture of SLSR at a low sampling rate.
- SLSR does not adopt complex topology information, such as, community, clique, and real global statistical characteristics of original networks, but can avoid the excessive preference for the core, and realize a balance sampling between core and periphery.